\documentclass[twocolumn]{aastex631}
\usepackage{graphicx,natbib,color,bm,url,times, hyperref}
\usepackage{amsmath}
\usepackage{amssymb}
\usepackage{euscript}
\usepackage{float}
\usepackage{capt-of}

\newcommand{\Figs}[2]{Figs~\ref{#1} and \ref{#2}}

\usepackage{tabularx}
\usepackage{array}
\usepackage{orcidlink}

\usepackage{hyperref}
\hypersetup{breaklinks=true, colorlinks=true, citecolor=blue}
\usepackage{color}

\usepackage{xcolor}

\begin{document}

\title{Effects of Dynamo-Generated Large-Scale Magnetic Fields on the Surface Gravity ($f$) Mode}

\author{Rajesh Mondal\orcidlink{0009-0005-7378-8614}}
\email{rajesh.mondal@iucaa.in}
\affiliation{Inter-University Centre for Astronomy \& Astrophysics, Post Bag 4, Ganeshkhind, Pune 411007, India}

\author{Nishant K. Singh\orcidlink{0000-0001-6097-688X}}
\email{nishant@iucaa.in}
\affiliation{Inter-University Centre for Astronomy \& Astrophysics, Post Bag 4, Ganeshkhind, Pune 411007, India}

\begin{abstract}
By modeling the upper layers of the Sun in terms of a two-layer setup where a free surface exists within the computational domain, we numerically study the interaction between the surface gravity, or the fundamental ($f$) mode, and the magnetic fields. Earlier such works were idealized in the sense that the static magnetic fields were imposed below the photosphere, i.e., the free surface, to detect signatures of subsurface magnetic fields and flows on the $f$-mode. In this work, we perform three-dimensional (3D) numerical simulations where the interior fluid below the photosphere is stirred helically at small scales, thus facilitating an $\alpha^2$ dynamo. This allows us to investigate how these self-consistently generated large-scale magnetic fields influence the properties of the $f$-mode. We find that when the magnetic fields saturate near the equipartition values with the turbulent kinetic energy of the flow, the $f$-mode is significantly perturbed. Compared to the nonmagnetic case, or the kinematic phase of the dynamo when fields are too weak, we note that the frequencies and the strengths of the $f$-mode are enhanced in the presence of saturated magnetic fields, with these effects being larger at larger wavenumbers. This qualitatively confirms the earlier findings from observational and numerical works, which reported the $f$-mode strengthening due to strong subsurface magnetic fields.

\end{abstract}

\section{Introduction} \label{sec:intro}

Techniques of helioseismology, especially those involving the acoustic modes
or the $p$-modes, have been extremely useful in measuring the interior properties of
the Sun \citep{RevModPhys.74.1073,2005LRSP....2....6G}.
The Sun also supports the fundamental mode, or the $f$-mode, whose eigenfunction lies only
close to the photosphere. Due to its simplistic dispersion relation, which is not sensitive
to the background thermodynamics, the $f$-mode is often considered to be of lesser
diagnostic value. However, a number of earlier works suggest that the $f$-modes can
be significantly perturbed in the presence of the magnetic fields
\citep{1961hhs..book.....C,1981SoPh...69...27R,Miles1989, Miles1992, Miles1992b,1997ApJ...486L..67C,2005A&A...431.1083E,Hanasoge_2008,
2009ApJ...694..573P,2014ApJ...795L...8S,10.1093/mnras/stu2540,
2017ApJ...846..162K,2022ApJ...934...61T}.

In addition to these studies, several observational works have explored the use of the
$f$-mode as a diagnostic of subsurface magnetic fields. For example, \citet{Dziembowski_2005} analyzed frequency shifts of both $p$- and $f$-modes and inferred the presence of subsurface magnetic fields with strengths of order $500\,{\rm G}$ at depths of approximately $5\,{\rm Mm}$; see also the review by \cite{Thompson_2006}.
\citet{2016SPD....47.0711S} reported a systematic strengthening of the $f$-mode
1-2 days before the emergence of active regions (ARs), based on their analysis
of the data from the Helioseismic and Magnetic Imager on board the
Solar Dynamics Observatory.
Their findings were later confirmed by \citet{waidele2023strengthening}
using a different technique involving the Fourier–Hankel analysis.
These studies signify the importance of the $f$-mode as a useful tool in
predicting the emergence of sunspots or ARs, and this topic
remains an area of active research; see also \citet{2022A&A...665A.141K, 2024arXiv240917421K, 2026arXiv260113145K}.

Properties of the $f$-mode in the presence of the magnetic fields have also been
studied extensively in various numerical works.
For example, \citet{Daiffallah2011, 2012ApJ...757..148F} studied the scattering of $f$-modes by thin magnetic flux tubes and found that the amplitude of the scattered wave approximately scales with the magnetic flux of the flux tubes.
\citet{10.1093/mnras/stu2540} through their piecewise isothermal setups in 2D
studied the properties of $p$- and $f$-modes by considering a variety of magnetic
configurations. When nonuniform magnetic fields were used in a similar setup,
it was found that the $f$-mode fans out in the diagnostic $k\omega$ diagrams,
and it displays strengthening as compared to the $f$-mode strengths obtained in the
purely hydrodynamic cases \citep{2014ApJ...795L...8S,2020GApFD.114..196S}.
They also demonstrated that the $f$-mode amplitude sensitively depends on the
depth of magnetic field concentrations, further highlighting its potential for
mapping subsurface magnetic fields.

Extending these works further, \citet{2024arXiv240914840K} performed more realistic convection simulations in 3D, and in qualitative agreement with earlier reports on the $f$-mode strengthening, they found that when strong magnetic fields are imposed just beneath the photosphere, the $f$-mode amplitudes become significantly larger compared to the cases with either weak or no magnetic fields. What is common in all these numerical works is that the magnetic fields are imposed, and their strengths are relatively strong compared to the local equipartition values.

In this work, we relax this assumption of an artificially imposed magnetic field by exciting an $\alpha^2$ dynamo \citep{BRANDENBURG20051} through a helical forcing of the fluid below the free surface. This allows us to study the interaction between the $f$-mode and a self-consistently generated equipartition-level large-scale magnetic field that is produced by the helical forcing in the bulk. As the magnetic fields are inherently buoyant, we do expect that the fields thus produced in this work will
also exist near the photosphere at similar equipartition levels.

In Section~\ref{sec:2}, we describe the model and numerical setup used in this study. Section~\ref{sec:3} outlines the methodology for analyzing the $f$-mode.
In Section \ref{sec:4}, we present our results by exploring the effects of self-sustained large-scale magnetic fields on the $f$-mode. We conclude in
Section~\ref{sec:5}.

\section{Model and Numerical Setup}\label{sec:2}
\subsection{Model}

We aim to investigate the influence of dynamo-generated large-scale magnetic fields on high-degree $f$-modes, which remain confined near the solar surface and are therefore sensitive to subsurface magnetic fields. For this purpose, it is sufficient to consider a Cartesian box filled with conducting fluid, which can be regarded as a local representation of a small region of the solar surface. This approach allows us to avoid the numerical complexity of a spherical domain and focus on the underlying physical processes. 

\begin{figure}[ht!]
  \centering
  \vspace{-3mm}
  \includegraphics[width=\linewidth]{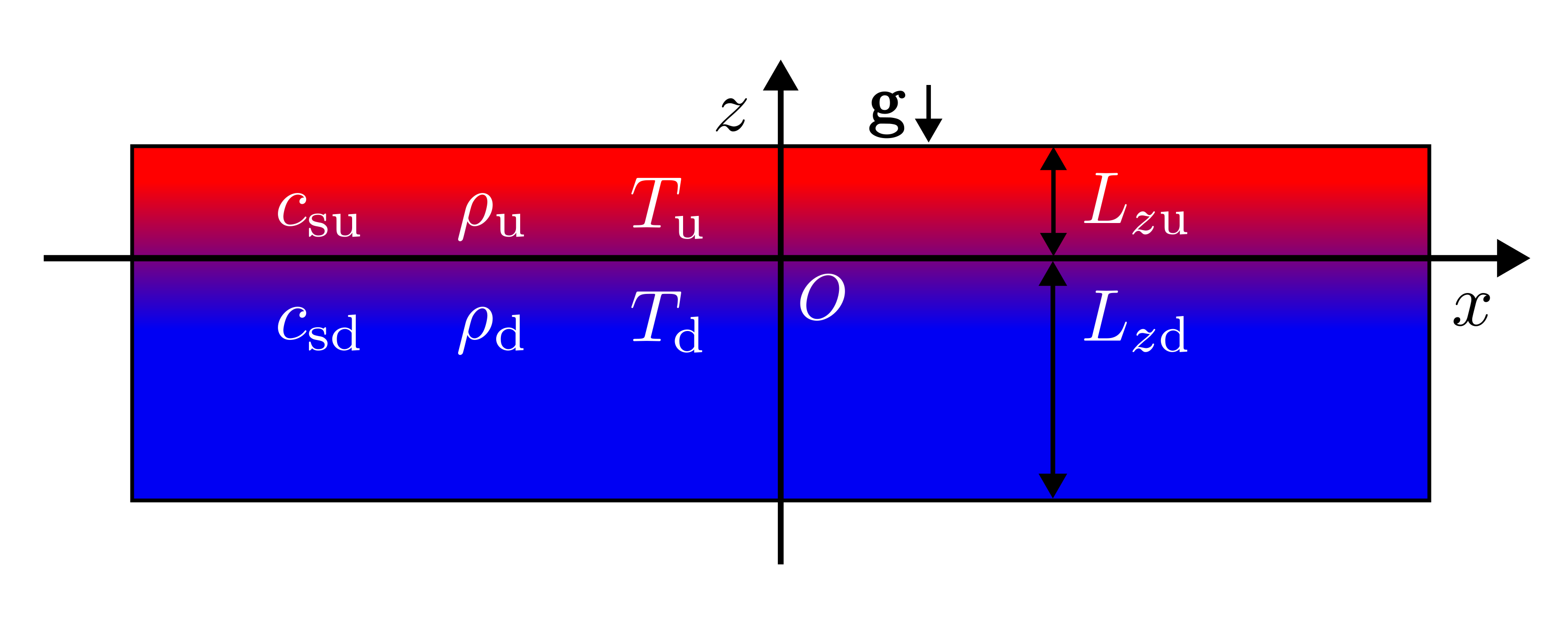}\\[1ex]
  \begin{minipage}{\linewidth}
  \includegraphics[width=0.94\linewidth]{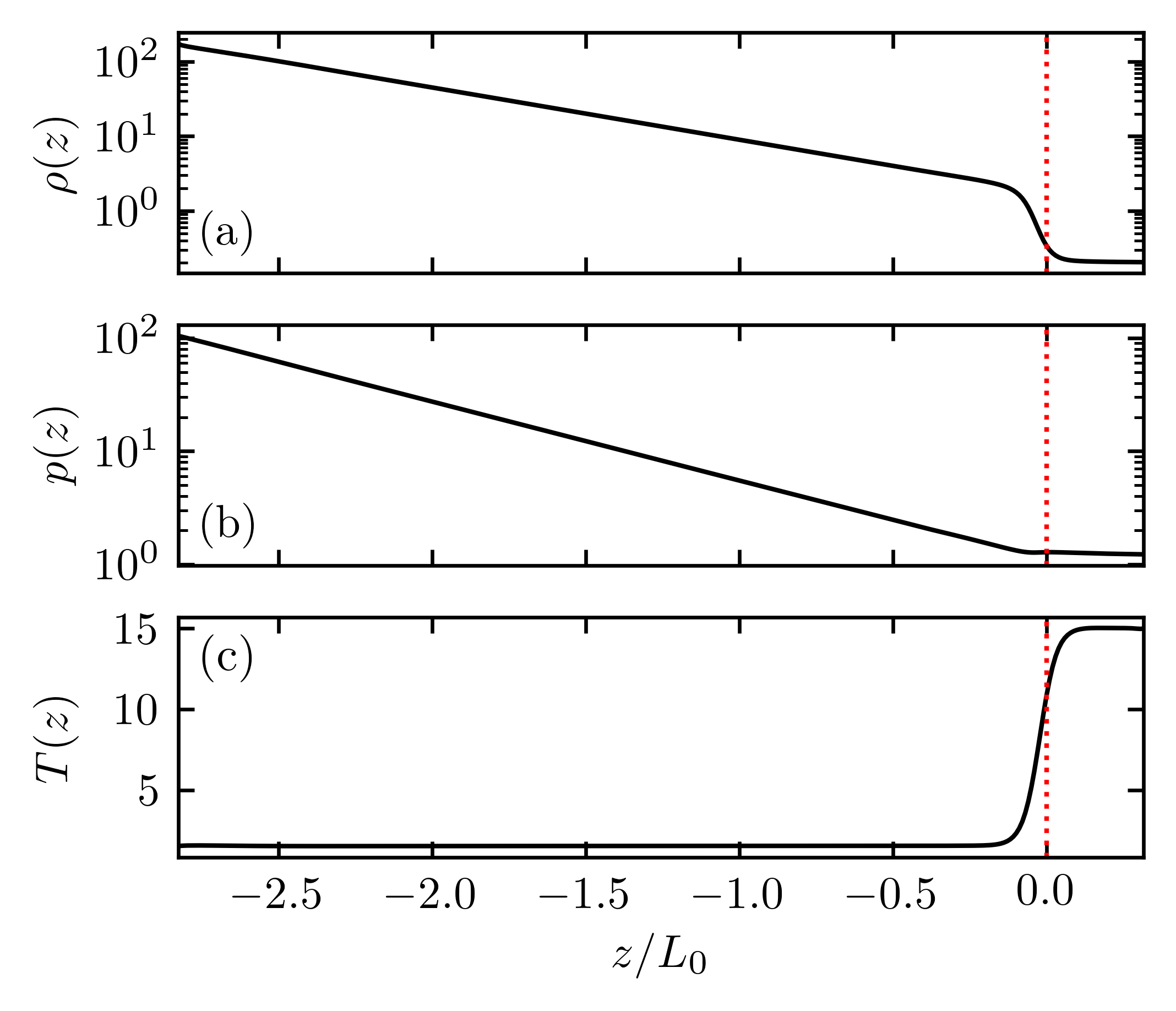}\hspace{0.5em}
  \end{minipage}
  \caption{
  Top: schematic of the two-layer simulation domain. Bottom: equilibrium profiles of (a) density, (b) pressure, and (c) temperature as a function of $z$ for $L_z/L_0 = \pi$. The red dotted line marks the interface at $z=0$.}
  \label{fig:1}
\end{figure}

We consider a Cartesian coordinate system with unit vectors $\mathbf{e}_x$, $\mathbf{e}_y$, and $\mathbf{e}_z$ along the $x$, $y$, and $z$ direction, respectively. The system consists of a Cartesian box of dimensions $L_x$, $L_y$, and $L_z$ along the $\mathbf{e}_x$, $\mathbf{e}_y$, and $\mathbf{e}_z$ directions, respectively, where $\mathbf{e}_z$ denotes the vertical direction (corresponding to the radial direction in the solar context) and $\mathbf{e}_x$, $\mathbf{e}_y$ the horizontal directions. For our purposes, we assume $L_z \ll L_x, L_y$, signifying that the system is significantly thinner in the vertical direction compared to the horizontal directions, so that we can work in the framework of parallel plane approximation. For reference, we have added a schematic diagram (2D cross-section) of the model, shown in the top panel of Figure~\ref{fig:1}.

Gravity acts along the $-\mathbf{e}_z$ direction, and its magnitude is denoted by $\mathrm{g}(>0)$. Given the small vertical extent of the domain, we have assumed gravity to be constant. This model also includes an interface located at $z=0$, which divides the domain into two subdomains, a lower layer of thickness $L_{z\mathrm{d}}$ and an upper layer of thickness $L_{z\mathrm{u}}$. We have assumed $L_{z\mathrm{d}}=9L_z/10$ and $L_{z\mathrm{u}}=L_z/10$ for all the simulations presented in this paper. 

We further assume a sharp discontinuity in the thermodynamic variables across the interface, as illustrated in the bottom panel of Figure~\ref{fig:1}. All the length scales are nondimensionalized using $L_0 = \gamma H_\mathrm{d} = c_\mathrm{sd}^2/\mathrm{g}$, where $H_\mathrm{d}$ is the isothermal pressure scale height of the lower subdomain and \(c_\mathrm{sd}\) is the adiabatic sound speed in the lower subdomain.

By assuming the conducting fluid to be ideal, we can write down the equation of state as $p=(c_\mathrm{p}-c_\mathrm{v})\rho T = \rho c_\mathrm{s}^2/\gamma$, where $p$, $\rho$ and $T$ are respectively the pressure, density and temperature, $c_\mathrm{p}$ and $c_\mathrm{v}$ are the specific heat at constant pressure and volume, $c_\mathrm{s}$ is the adiabatic sound speed and $\gamma=c_\mathrm{p}/c_\mathrm{v}$ is the adiabatic index. In this work, both layers are taken to be isothermal. While there are other possibilities, we have not explored them here. It would be interesting to investigate the effect of different kinds of layers and test the robustness of the results we discuss here. 

Since gravity is constant and both layers are isothermal, in hydrostatic equilibrium, both layers are exponentially stratified as
\begin{align}
    \rho_\mathrm{d,u}(z) = \rho_\mathrm{d,u}(0)\exp\left(-\frac{z}{H_\mathrm{d,u}}\right)
\end{align}
where $H_\mathrm{d,u} = (c_\mathrm{p}-c_\mathrm{v})T_\mathrm{d,u}/\mathrm{g}$ is the scale height of the lower and upper subdomains. Here, $\rho_\mathrm{d}(0)$ and $\rho_\mathrm{u}(0)$ are the densities at the interface, $T_\mathrm{d}$ and $T_\mathrm{u}$ ($>T_\mathrm{d}$) are the constant temperatures in the lower and upper subdomains, respectively. This defines our background state, on top of which perturbations are introduced.

This model can be regarded as a simplified representation of a small region of the Sun near its surface, with the interface mimicking the solar surface. The subdomain below the interface corresponds to the outer part of the convection zone, while the region above represents the solar atmosphere. This model, despite its simplicity, is sufficient for our purposes. A potential limitation is the use of an isothermal lower subdomain, whereas the solar convective zone is weakly superadiabatic. In future work, we will address this issue. For now, we adopt this simplified model, recognizing that it does not fully capture the thermodynamic structure of the solar interior.

\subsection{Evolution equations}

Having established the background state, we now describe the evolution equations used to excite the dynamo and the $f$-modes in our model.  We solve the following set of equations:

\begin{gather}
\frac{D~\mathrm{ln}\rho}{D t} = -\nabla\cdot\boldsymbol{u}\label{eq:3}\\
\frac{D\boldsymbol{u}}{D t} = \boldsymbol{f}+\boldsymbol{\mathrm{g}}+\frac{1}{\rho} \left(\boldsymbol{J}\times\boldsymbol{B} -\nabla{p}+\nabla\cdot2\nu\rho\boldsymbol{S} \right)\label{eq:4}\\
T\frac{Ds}{Dt} = 2\nu\boldsymbol{S}^2+\frac{\mu_0\eta}{\rho}\boldsymbol{J}^2-\gamma(c_\mathrm{p}-c_\mathrm{v})\frac{T-T_\mathrm{d,u}}{\tau_\mathrm{d,u}}\label{eq:5}\\
\frac{\partial \boldsymbol{A}}{\partial t} = \boldsymbol{u}\times\boldsymbol{B}-\mu_0\eta\boldsymbol{J}\label{eq:6}
\end{gather}

\noindent with appropriate boundary and initial conditions (discussed below). In the above equations, $\rho$ is the density, $\boldsymbol{u}$ is the velocity, and $D/Dt$ denotes the convective time derivative. The term $\boldsymbol{f}$ represents a forcing function, while $\boldsymbol{\mathrm{g}} = (0,0,-\mathrm{g})$ is the gravitational acceleration. The magnetic field is given by $\boldsymbol{B} = \nabla\times\boldsymbol{A}$, where $\boldsymbol{A}$ is the vector potential, $\boldsymbol{J}=\mu_0^{-1}(\nabla\times\boldsymbol{B})$ is the current density. The pressure is denoted by $p$, and the traceless rate-of-strain tensor is defined as
\begin{align}
S_{ij}=\frac{1}{2}(u_{i,j}+u_{j,i})-\frac{1}{3}\delta_{ij}\nabla\cdot\textbf{u},
\end{align}

\noindent where commas denote partial differentiation. The specific entropy is denoted by $s$, the temperature is denoted by $T$, and $\mu_0$ represents the vacuum permeability. The kinematic viscosity $\nu$ and magnetic diffusivity $\eta$ are taken to be constants. Finite values of $\nu$ and $\eta$ are included primarily to ensure numerical stability. These parameters are chosen to be sufficiently small so as not to affect the $f$-mode dynamics, while remaining large enough to suppress grid-scale instabilities. We find that values $\nu, \eta \sim 10^{-3}$–$10^{-4}$ are sufficient for our purposes.

The set of equations (\ref{eq:3})-(\ref{eq:6}) constitutes an almost standard set of MHD equations, extended to serve our specific purpose. The inclusion of the relaxation term in equation (\ref{eq:5}) can be motivated as follows. The presence of finite $\nu$ and $\eta$ in the evolution equations will lead to the conversion of kinetic and magnetic energy to thermal energy, which would otherwise heat the background state. To counteract this effect, we introduce a relaxation term that maintains the prescribed background stratification. This approach is motivated by \citet{10.1093/mnras/stu2540}. While they applied this term only in the upper subdomain, we find that in our 3D setup, it is necessary to include it in both subdomains. Without the relaxation term in the lower subdomain, the upper subdomain gradually heats the lower one, thereby reducing the sharpness of the interface. Since a well-defined interface is crucial for excitation of the $f$-mode, incorporating the relaxation term in both subdomains is essential. Thus, for the upper layer, we choose a relaxation rate $\tau_\mathrm{u}^{-1}=0.45~\mathrm{g}/c_{\mathrm{sd}}$, while for the lower layer we adopt $\tau_\mathrm{d}^{-1}=0.25~\mathrm{g}/c_{\mathrm{sd}}$ throughout this paper.

Our background state is static, so to introduce perturbations, we have included a stochastic forcing term (\cite{Brandenburg_2001}) in the equation (\ref{eq:4}). This term serves two purposes: it introduces small perturbations into the background state in an isotropic and homogeneous manner at a length scale smaller than the box size, thereby exciting various modes; second, the forcing is chosen to be helical, injecting kinetic helicity into the flow and breaking mirror symmetry, which is a necessary condition for large-scale dynamo action. Since we aim to excite the dynamo primarily in the lower layer, the forcing is set to zero in the upper layer.

The average forcing wavenumber $k_\mathrm{f}$ defines the energy injection scale $l_\mathrm{f} = 2\pi/k_\mathrm{f}$. The vertical box wavenumber is $k_z = 2\pi/L_z$, giving $k_\mathrm{f}/k_z \approx 4.75$ for the setup used here. We also note that $l_\mathrm{f}/L_0 \sim 0.7$, where $L_0$ is the adiabatic scale height defined above. In the mixing length \citep{2023Galax..11...75J}, the dominant energy-carrying scale at a given depth is expected to scale with the local pressure scale height. The chosen forcing scale is therefore motivated by this physical consideration.

Because the forcing scale in our model is relatively small, the Rossby number $\mathrm{Ro} = U/(\Omega L)$ would be large in the solar context, and the role of Coriolis-induced helicity at such scales is therefore uncertain. However, our goal is not to model the solar dynamo itself. Rather, the dynamo included in our simulations serves only as a mechanism to generate a near-equipartition large-scale magnetic field self-consistently. From mean-field dynamo theory we expect that an $\alpha^2$ dynamo generates magnetic fields at scales comparable to the system size, largely independent of the forcing scale
\citep{BRANDENBURG20051}. Also, adopting a relatively large $k_\mathrm{f}$ keeps the magnetic Reynolds number $\mathrm{Rm} = u_\mathrm{rms,d}/(\eta k_\mathrm{f})$ modest and thereby suppresses the excitation of a small-scale dynamo.

% The smallest wavenumber in the system is $k_1 = 2\pi/\max(L_x, L_y, L_z) = 2\pi/L_x$, which in our setup corresponds to
% $k_1 = 2\pi/L_x$. For the box size used in this work, we have $k_\mathrm{f}/k_1 \approx 56$. 

\subsection{Boundary conditions}

We apply periodic boundary conditions for all eight variables at the boundaries in the horizontal direction. However, in the stratified vertical $z$ direction, we impose physically motivated boundary conditions at the top and bottom. For velocity, we choose the free slip condition to avoid the emergence of boundary layers, which are unimportant in an astrophysical context. For the entropy equation, we fix the temperature at both boundaries, while for density, we set its second derivative to zero at the boundaries. For the magnetic field, we apply perfectly conducting boundary conditions at both the top and bottom.

\subsection{Initial conditions}

We set the initial velocity to zero throughout the domain. For the density, we prescribe the two-layer profile shown in panel (a) of Figure~\ref{fig:1} (bottom). Using density and temperature profiles together with the fundamental thermodynamic relations, we compute the corresponding entropy profile for our background state and prescribe it as the initial condition for the entropy in a manner analogous to density. Finally, for the magnetic field, we assume a normally distributed white noise of small amplitude as the initial seed magnetic field.

\subsection{Software}

All simulations are performed using the \textsc{Pencil Code} \footnote{\href{https://pencil-code.nordita.org/}{https://pencil-code.nordita.org/}} \citep{2021JOSS....6.2807P}, a publicly available high-order finite-difference code for weakly compressible magnetohydrodynamics. 

All the analyses are carried out using publicly available Python packages, including \textsc{matplotlib} \citep{2007CSE.....9...90H},
\textsc{NumPy} \citep{2011CSE....13b..22V},
\textsc{SciPy} \citep{Virtanen2020} and \textsc{h5py} \citep{2023zndo...7560547C}.

\section{Methodology}\label{sec:3}

To investigate the effect of the magnetic field on $f$-modes, we use one of the fundamental tools of helioseismology \citep{2005LRSP....2....6G}: $k$-$\omega$ diagram. For our setup, calculating the $k$-$\omega$ diagram from simulated data is straightforward and analogous to its observational counterpart (essentially the same). We save 2D snapshots of the vertical velocity component $u_z(t, x, y, z=0)$ at the interface ($z=0$) at regular time intervals, analogous to dopplergrams in observations. The snapshots are stacked sequentially to form a 3D data cube. We then perform the Fourier transform along all three axes to obtain $\hat{u}_z(\omega, k_x, k_y)$. The corresponding energy spectral density is given by $P(\omega, k_x, k_y) = \left|\hat{u}_z\right|^2$. Finally, to construct the $k$-$\omega$ diagram, we may choose either $k_y=0$ or $k_x=0$. In all results presented below, we adopt $k_y = 0$. 

The results are presented in terms of dimensionless quantities. This is not strictly necessary for the present study since all the relevant physical parameters are identical for all the runs. Still, we have adopted this normalization for consistency and to facilitate potential future follow-up work.  Specifically, we use the same normalization as discussed in \citet{10.1093/mnras/stu2540}:
\begin{equation}
    \tilde{k}_x = k_x L_0, \quad \tilde{\omega} = \frac{\omega}{\omega_0}, 
\end{equation}
where $L_0 = \gamma H_\mathrm{d} = c_\mathrm{sd}^2/\mathrm{g}$ and $\omega_0 = \mathrm{g}/c_\mathrm{sd}$ represents the inverse of the characteristic timescale for acoustic wave propagation in a stratified medium. Remembering that $\hat{u}_z$ has the dimension of $[LT^{-1}]$, we nondimensionalize the power spectrum as  
\begin{equation}
    \tilde{P} = \frac{P}{c_{\mathrm{sd}}^2} = \left(\frac{|\hat{u}_z|}{c_{\mathrm{sd}}}\right)^2 = \left(\frac{|\hat{u}_z|}{L_0\omega_0}\right)^2.
\end{equation}
where, $c_{\mathrm{sd}}$ is the adiabatic sound speed of the lower subdomain. Thus, $\tilde{P}$ is simply the squared Mach number of the vertical velocity component in Fourier space.

\subsection{Mode fitting}\label{sec:3.1}

To quantify the behavior of the $f$-mode, we extract the power spectrum $\tilde P_{\tilde k_x}(\tilde \omega)$ as a function of $\tilde \omega$ for different $\tilde k_x$. Each profile is then fitted using a Lorentzian function to represent the mode and a linear function to model the background continuum. Therefore, the fitting function we use is of the following form:

\begin{equation}
    \tilde P_{\tilde k_x}(\tilde \omega) = \sum_{i}\frac{A_i~\Gamma_i/\pi}{(\tilde \omega-\tilde \omega_{\mathrm{c},i})^2+\Gamma_i^2} + B~\tilde \omega + C
\end{equation}

where $A_i$, $\tilde \omega_{\mathrm{c},i}$, and $\Gamma_i$ denote the amplitude, central frequency, and linewidth of the $i$th mode, respectively, while B and C characterize the background. The choice of a linear background is not motivated by any theoretical reasoning but rather by empirical observations from the dataset. For the fitting procedure, we employ SciPy's \texttt{curve\_fit} function, which provides reliable and accurate results. 

For low $k_x$, where $f$- and $p$-modes are closely spaced in frequency (see Figure~\ref{fig:2} and Figure~\ref{fig:4}), we therefore fit them simultaneously assuming a Lorentzian profile for each of them. At larger $k_x$, where modes are well separated, only the $f$-mode and the background are fitted.

To characterize the $f$-mode, we adopt three diagnostic parameters proposed by \cite{10.1093/mnras/stu2540}, evaluated at fixed $\tilde k_x$ from the Lorentzian profile obtained after subtracting the linear background from $\tilde P_{\tilde k_x}(\tilde \omega)$. The first is the relative frequency shift, defined as

\begin{equation} \left(\frac{\delta\omega_\mathrm{f}^2}{\omega_\mathrm{f}^2}\right)(\tilde k_x) = \frac{\omega_\mathrm{cf}^2-\omega_\mathrm{f}^2}{\omega_\mathrm{f}^2}, \end{equation}

where $\omega_\mathrm{f}$ is the theoretical frequency from equation (\ref{eq:14}), and $\omega_\mathrm{cf}$ is the fitted line-center frequency of the $f$-mode. This parameter quantifies the shift of the line center relative to its theoretical value.

To quantify the amplitude of the $f$-mode, we define mode strength $\mu_\mathrm{f}$ as the integrated excess energy in the mode frequency range:

% \begin{equation} \mu_\mathrm{f}(k_x) = \int\Delta\left|\hat{u}_z\right|^2\hspace{0.5mm}d\omega, \end{equation}
\begin{equation} \mu_\mathrm{f}(\tilde k_x) = \int\Delta\tilde P_{\tilde k_x}(\tilde \omega)\hspace{0.5mm}d\omega, \end{equation}

where $\Delta\tilde P_{\tilde k_x}(\tilde \omega)$ represents the normalized energy excess over the continuum of the $f$-mode. This quantity serves as a proxy for the kinetic energy of the mode.

Finally, the relative line width is quantified using the dimensionless parameter:

\begin{equation} \Gamma_\mathrm{f}(\tilde k_x) = \frac{\Delta\omega_\mathrm{FWHM}}{\omega_\mathrm{cf}}, \end{equation}

where $\Delta\omega_\mathrm{FWHM}$ is the full width at half-maximum (FWHM) of the fitted line profile. Note that this relative line width is the inverse of the standard quality factor, $Q$.

\section{Results}\label{sec:4}

To study how a dynamo-generated large-scale magnetic field affects the $f$-mode, we perform two simulations in total. One of them, labeled h1, is a purely hydrodynamic case. The other run, d1, is hydromagnetic.
Both simulations are carried out at the same grid resolution of $1024 \times 256 \times 320$, with identical physical parameters across the runs. The computational domain has dimensions of $12\pi \times 3\pi \times \pi$ (in units of $L_0$) and a constant kinematic viscosity $\nu = 0.001$. For the magnetic case (d1), we additionally employ a constant magnetic diffusivity $\eta = 0.001$. These choices correspond to an approximate Reynolds number of $\mathrm{Re} \approx 18$, and a magnetic Prandtl number $\mathrm{Pm} = 1$ in the hydromagnetic simulations.

\subsection{Nonmagnetic case}

\begin{figure}[H]
\centering
  \includegraphics[width=\columnwidth]{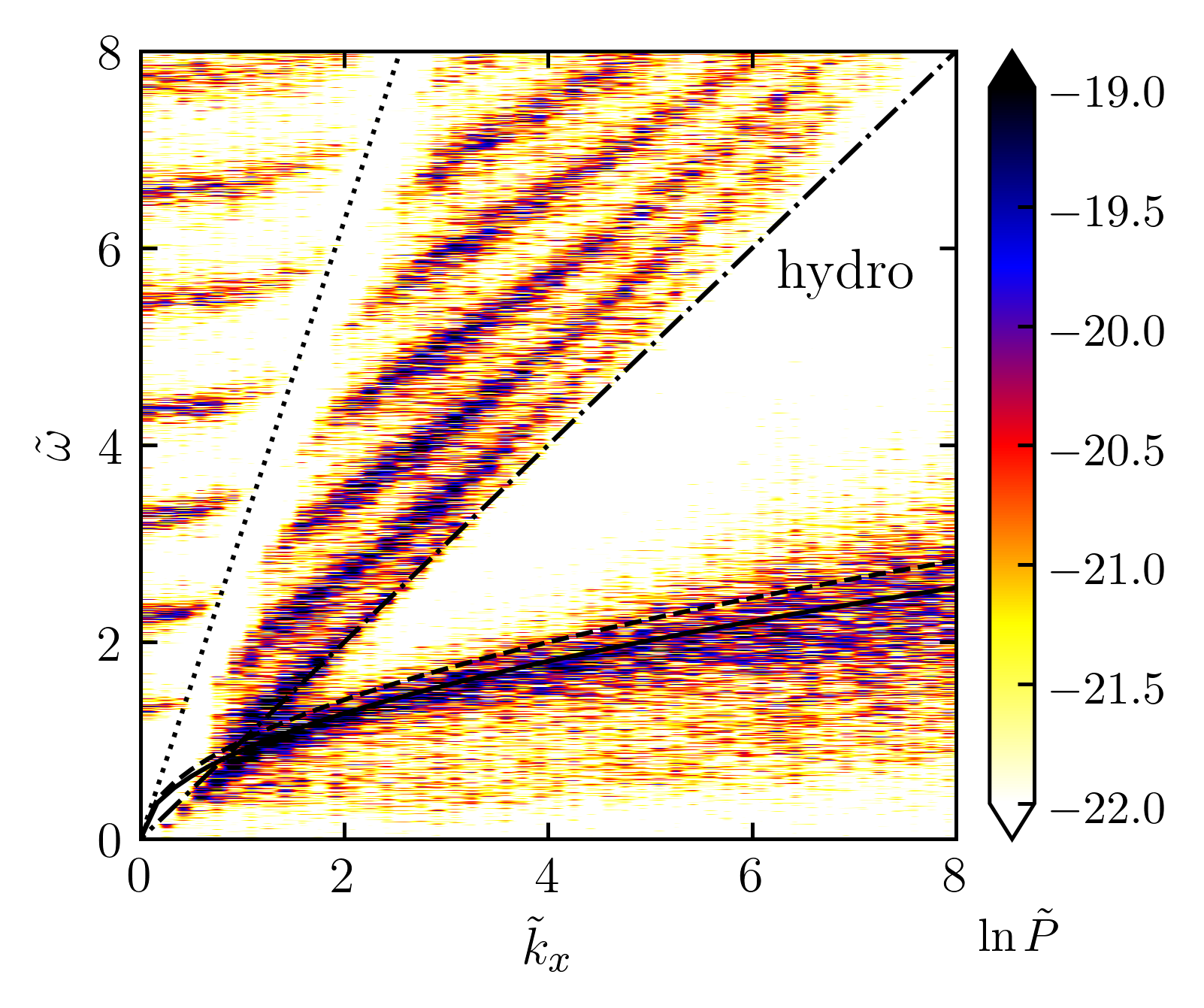}
  \caption{$k$-$\omega$ diagram for the nonmagnetic run h1. The dotted and dotted-dashed lines show $\tilde\omega=c_\mathrm{su}\tilde k_x$ and  $\tilde\omega=c_\mathrm{sd}\tilde k_x$ respectively. The dashed and solid curves show $\tilde\omega_\mathrm{f0}$ and $\tilde\omega_\mathrm{f}$, respectively.}
  \label{fig:2}
\end{figure}

Before turning to the magnetic case, we first discuss the wave modes in the purely hydrodynamic run h1. The $k$-$\omega$ diagram for this run is shown in Figure~\ref{fig:2}. The $k$-$\omega$ diagram reveals the excitation of both the $f$-mode and several $p$-modes in the domain. For a two-layer isothermal setup like ours, four critical lines exist in the $k$-$\omega$ diagram. Of these, we show two: $\omega = c_\mathrm{su} k_h$ (dashed line) and $\omega = c_\mathrm{sd} k_h$ (solid line). The remaining two critical lines (not shown) are $\omega=N_\mathrm{u}$ and $\omega=N_\mathrm{d}$, where $N_\mathrm{u}$ and $N_\mathrm{d}$ are the \emph{Brunt-Väisälä} \citep{1974ARA&A..12..407S} frequencies of the upper and lower layers, respectively. 

Since our primary interest in this work is the $f$-mode, we briefly summarize its theoretical properties. The $f$-mode is a surface gravity wave that exists due to a discontinuity in the density profile. For our model, the dispersion relation of the $f$-mode can be calculated analytically \citep{1932hydr.book.....L} and is given by
\begin{equation}
    \omega_\mathrm{f}^2 = \omega_\mathrm{f0}^2 \frac{1 - q}{1 + q}, 
    \quad 
    \omega_\mathrm{f0}^2 = \mathrm{g} k_h ,
    \label{eq:14}
\end{equation}
where $k_h$ is the horizontal wavenumber; for the present purposes, $k_h = k_x$. The parameter $q$ is defined as
\begin{equation}
    q = \frac{\rho_\mathrm{u}(0)}{\rho_\mathrm{d}(0)}
      = \frac{c_\mathrm{sd}^2}{c_\mathrm{su}^2}
      = \frac{H_\mathrm{d}}{H_\mathrm{u}}
      = \frac{T_\mathrm{d}}{T_\mathrm{u}},
\end{equation}
and characterizes the jump in the thermodynamic variables across the interface at $z = 0$. Here, $\omega_\mathrm{f0}$ denotes the angular frequency of the $f$-mode in the limit $q \to 0$. The dashed and solid curves in Figure~\ref{fig:2} show $\omega_\mathrm{f0}$ and $\omega_\mathrm{f}$, respectively.

For the low values of $\tilde{k}_x$, $f$-mode frequencies lie close to the $\omega_\mathrm{f}$ curve, whereas for large $\tilde k_x$, they deviate from it and the mode also broadens. Such broadening can be attributed to the turbulent nature of the flow \citep{1999A&A...349..312M,2000A&A...360..707M}.

To facilitate comparison, we present the variation of all three mode parameters (discussed above) in the next section (see Figure~\ref{fig:5}) alongside the corresponding values obtained from the hydromagnetic run.

\subsection{With \texorpdfstring{$\alpha^2$}{alpha^2} dynamo}
\begin{figure*}[htb!]
  \centering
  \begin{minipage}[c]{0.48\textwidth}
    \centering
    \raisebox{1mm}{\includegraphics[width=\linewidth]{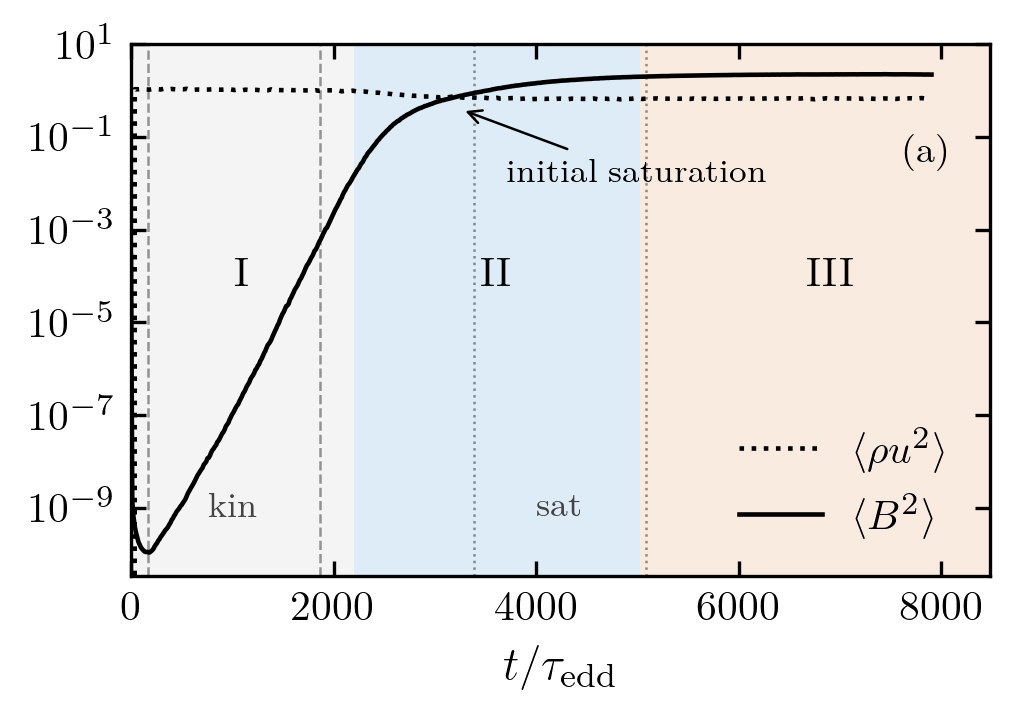}}
    \label{fig:3a}
  \end{minipage}
  \hfill
  \begin{minipage}[c]{0.48\textwidth}
    \centering
    \raisebox{0.95mm}{\includegraphics[width=\linewidth]{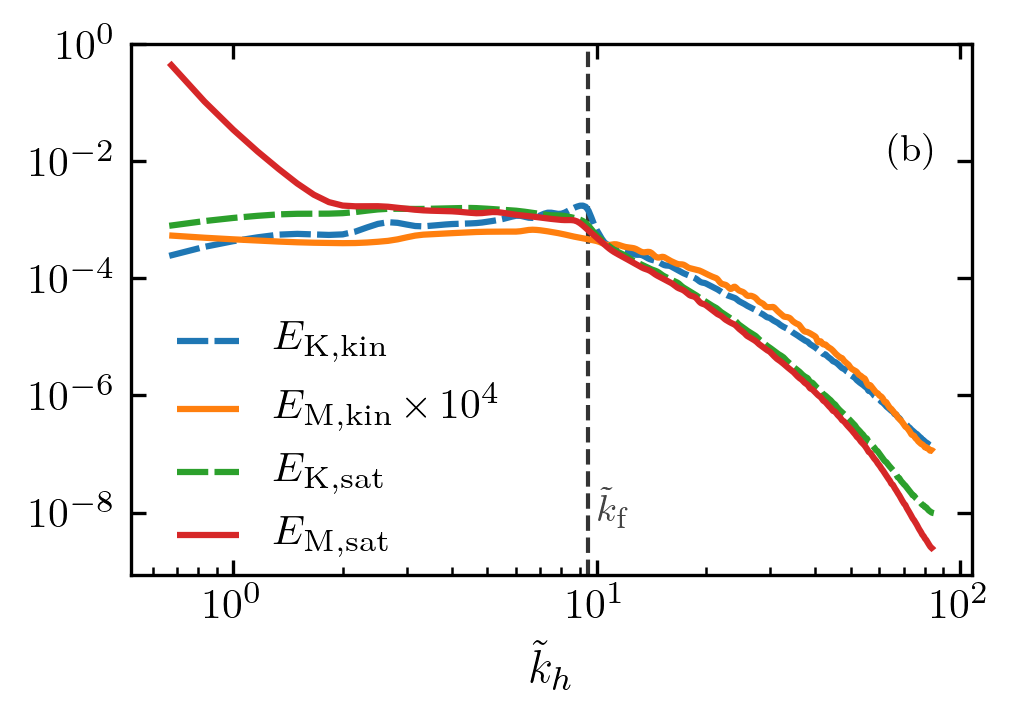}}
    \label{fig:3b}
  \end{minipage}
  % \vspace{0.5mm}  % spacing between rows
  % \vspace{-2.5cm}  % spacing between rows

  % bottom row: asymmetric split
    \begin{minipage}[c]{0.71\textwidth}
      \centering
      \raisebox{1.75mm}{\includegraphics[width=\linewidth]{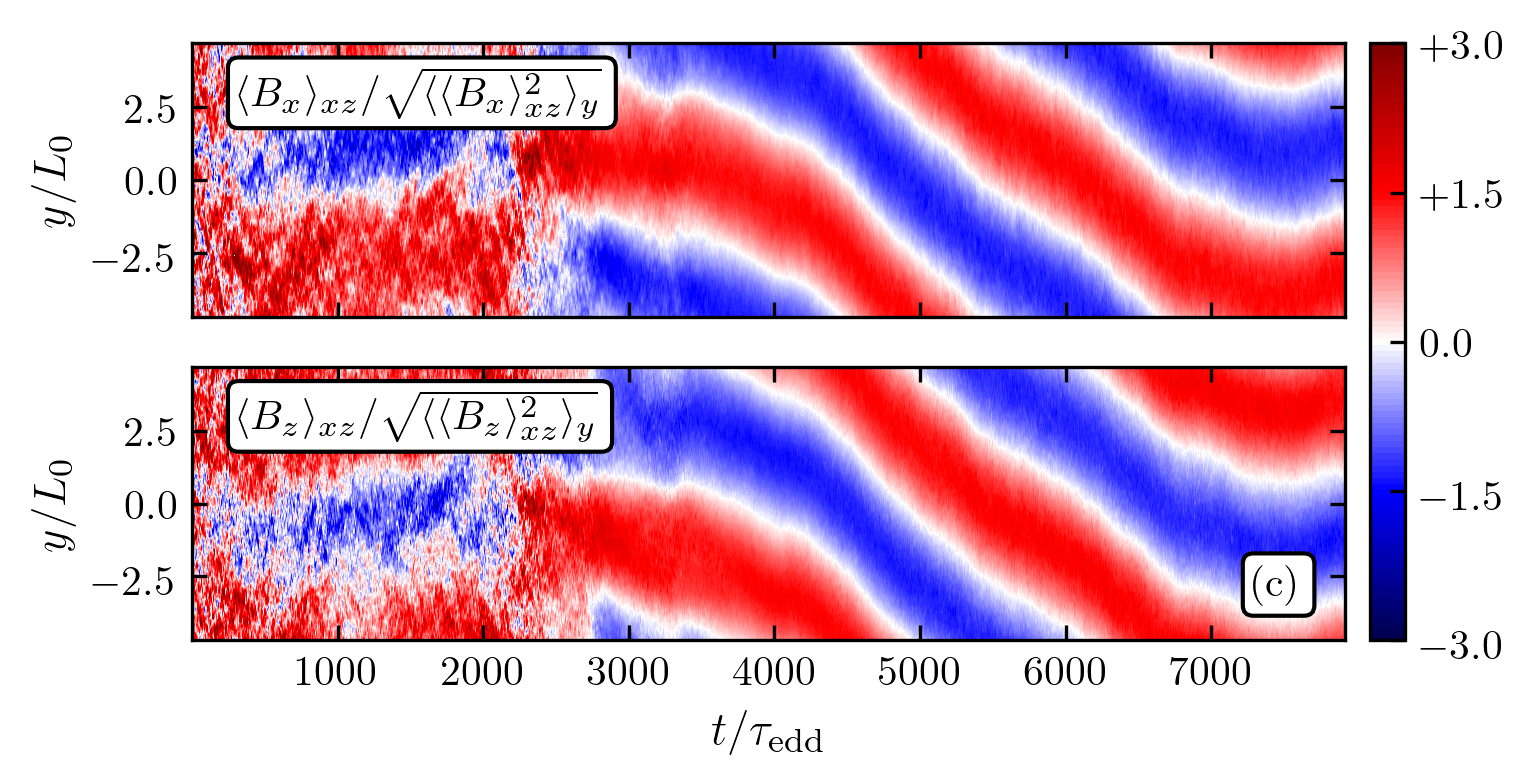}}
      \label{fig:3c}
    \end{minipage}
    \hfill
    \begin{minipage}[c]{0.268\textwidth}
      \centering
      \makebox[\linewidth][l]{%
    \includegraphics[width=0.975\linewidth]{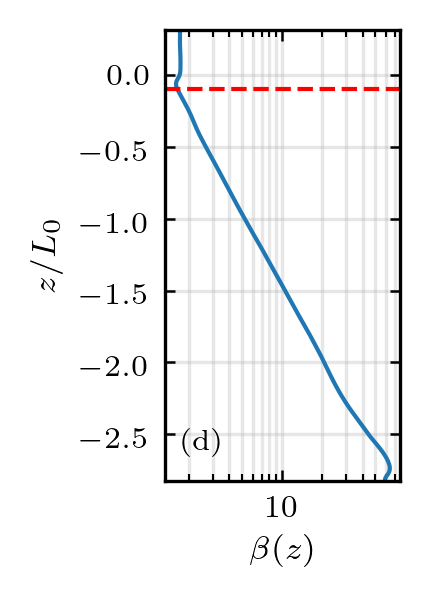}
     }
      \label{fig:3d}
    \end{minipage}
  \caption{(a) Time evolution of the kinetic and magnetic energies for run d1. Three phases, (I) kinematic, (II) weak nonlinear, and (III) saturated, of the $\alpha^2$ dynamo are highlighted. (b) Kinetic and magnetic energy spectra are shown from the plane at $z=-0.1L_0$, during the kinematic and saturated phases. The black dashed line shows the forcing wavenumber $k_\mathrm{f}$. (c) Spacetime diagram of the mean magnetic field components. Magnetic fields are first averaged over the $x$-$z$ plane to define $\langle B_x\rangle_{xz}$ and $\langle B_z\rangle_{xz}$. These quantities are then normalized by their corresponding rms values $\sqrt{\langle\langle B_x\rangle_{xz}^2\rangle_y}$ and $\sqrt{\langle\langle B_z\rangle_{xz}^2\rangle_y}$, respectively. (d) Vertical profile of horizontally averaged plasma beta $\beta(z)$, in the saturated phase. 
  % Black circles show the eigenfunction of the $f$-mode near the free surface, scaled by a factor of $10^{10}$ for clarity. 
  The red dashed line marks $z=-0.1L_0$.}
  \label{fig:3}
\end{figure*}

\begin{figure*}
  \centering
  \begin{minipage}[c]{0.48\textwidth}
    \centering
    \includegraphics[width=\linewidth]{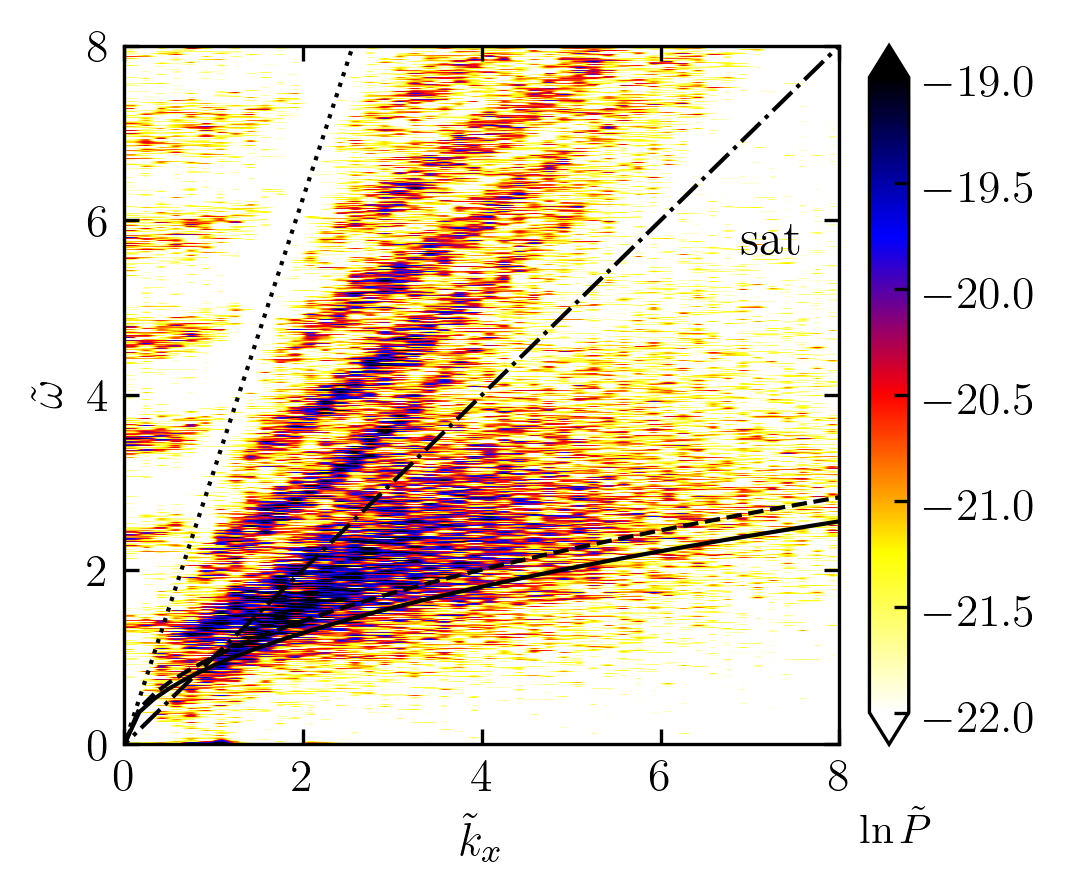}
  \end{minipage}
  \hfill
  \begin{minipage}[c]{0.48\textwidth}
    \centering
    \includegraphics[width=\linewidth]{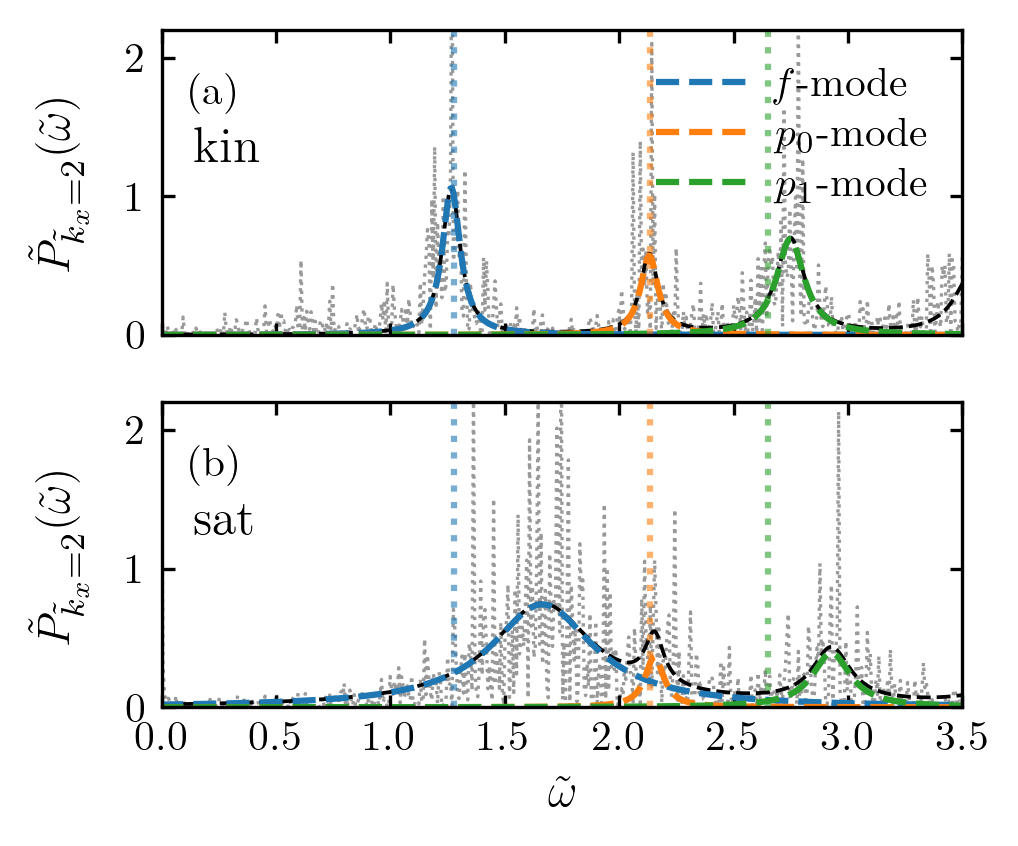}
  \end{minipage}
  \caption{
    Left: same as Figure~\ref{fig:2} but for the magnetic run d1. A noticeable broadening of the $f$-mode is seen with increasing $\tilde k_x$. Right: line profiles of the $f$-, $p_0$-, and $p_1$-mode at $\tilde{k}_x=2$ for the kinematic phase (a) and for the saturated phase (b); dotted curves represent the data. To ensure a consistent comparison, the background has been removed. Blue dashed lines show the locations of the $f$-mode.
  }
  \label{fig:4}
\end{figure*}

We begin by examining the evolution of the magnetic field itself, before turning to the behavior of the $f$-mode in the presence of magnetic fields. As mentioned above, we incorporate an $\alpha^2$ dynamo \citep{BRANDENBURG20051} in the lower subdomain, where an initial weak seed magnetic field is amplified through the dynamo action. Figure~\ref{fig:3}a shows the time evolution of the kinetic and magnetic energies for the hydromagnetic run d1. Time is expressed in units of eddy turnover time $\tau_\mathrm{edd}=1/(u_\mathrm{rms,d}\,k_\mathrm{f})$, where $u_\mathrm{rms,d}$ is the rms velocity measured in the lower layer. Since the dynamo operates only in the lower layer, this definition appropriately characterizes the relevant dynamical timescale. The growth of the magnetic energy exhibits an initial exponential phase, indicative of the kinematic dynamo regime, followed by a transitional phase and eventual saturation at later times. In the saturated phase, the magnetic field reaches an rms strength of $B_\mathrm{rms}=1.4~B_\mathrm{eq}$, where the equipartition field is defined as $B_\mathrm{eq}=(\mu_0\rho_\mathrm{d} u_\mathrm{rms,d}^2)^{1/2}$.

We select two representative time intervals, one from the kinematic phase and another from the postequipartition stage, to examine how a dynamically evolving magnetic field influences the $f$-mode. These intervals are marked in the Figure~\ref{fig:3}a. For the kinematic phase, we consider the data between the dashed vertical lines (hereafter \emph{kin}), while for the postequipartition stage (hereafter \emph{sat}), we consider the data between the dotted vertical lines.

The Figure~\ref{fig:3}b shows various 2D energy spectra at \(z = -0.1~L_0\), which is the same depth used to construct the \(k\)–\(\omega\) diagrams analyzed below. Although the kinetic energy spectrum for the nonmagnetic run h1 is not shown here, it is identical to \(E_\mathrm{K,kin}\), as expected. Two main observations can be made from this figure. 

First, the excitation of the \(\alpha^2\) dynamo is evident. We find that the magnetic energy in the kinematic phase (\(E_\mathrm{M,kin}\)) is approximately \(10^{-4}\) times smaller than that in the saturated phase (\(E_\mathrm{M,sat}\)). Moreover, \(E_\mathrm{M,sat}\) peaks near the largest scale of the box, as expected. 

Second, we observe that the kinetic energy shifts from smaller scales to larger scales as the dynamo saturates. This increase in kinetic energy at large scales can, in turn, strengthen different modes. Below, we confirm that this is indeed the case.

Figure~\ref{fig:3}c shows the spacetime diagrams of the different mean magnetic field components for run d1, illustrating the evolution of the magnetic field structure in our box. The field evolves from an initially noisy state to large-scale dynamo waves, exhibiting typical \(\alpha^2\) dynamo behavior.

% Figure~\ref{fig:3}d shows the horizontally averaged plasma beta $\beta(z)$ for the saturated phase, with black circles indicating the $f$-mode eigenfunction near the interface ($z=0$). In our setup, the $f$-mode decays exponentially on both sides of $z=0$. The small values of $\beta$ near the interface suggest that magnetic pressure dominates within the $f$-mode cavity.

% \blue{Figure~\ref{fig:3}d shows the horizontally averaged plasma beta $\beta(z)$ for the saturated phase. 
% The profile illustrates the vertical variation of the ratio of gas pressure to magnetic pressure in the simulation domain. 
% Near the interface ($z=0$), the relatively small values of $\beta$ indicate that magnetic pressure becomes comparable to or exceeds the gas pressure in this region. This is expected since pressure is falling above the interface therefore magnetic pressure is relatively dominating above the interface, not because we have more magnetic field there.} 

% \blue{In setups of this kind, the $f$-mode eigenfunction is known to decay exponentially away from the interface; see Singh et al.~(2015) for an illustration of the qualitative structure of the $f$-mode eigenfunction in a similar configuration.}

Figure~\ref{fig:3}d shows the horizontally averaged plasma beta $\beta(z)$ in the saturated phase, illustrating the vertical variation of the ratio of gas pressure to magnetic pressure in the simulation domain. Near the interface ($z=0$), the relatively small values of $\beta$ indicate that the magnetic pressure becomes comparable to, or exceeds, the gas pressure in this region. This behavior arises primarily because the gas pressure decreases rapidly above the interface, rather than because the magnetic field strength is significantly larger there.

% Consequently, the region of low $\beta$ near the surface should not be interpreted as representing sunspot-like magnetic fields, but rather as a consequence of the imposed stratification of the atmosphere.

In configurations of this type, the $f$-mode eigenfunction is known to decay exponentially away from the interface; see Singh et al.~(2015) for an illustration of the qualitative structure of the $f$-mode eigenfunction in a similar setup.

Having established the presence and the evolution of the dynamo in our simulation, we now turn to the wave dynamics of the system, focusing on the $k$-$\omega$ diagram and the $f$-mode. Our approach is straightforward: by analyzing the power spectrum of vertical velocity fluctuations at the interface, we identify the characteristic signatures of different wave modes and assess how they are influenced by the dynamo-generated magnetic fields. Figure~\ref{fig:4} (left) shows the $k$-$\omega$ diagram for the saturated phase of the run d1; the $k$-$\omega$ diagram for the kinematic phase is not shown here, but it is similar to the nonmagnetic case (Figure~\ref{fig:2}). We use the same data length for the kinematic and saturated intervals marked in Figure~\ref{fig:3}a to ensure a consistent comparison.

Examining the $k$-$\omega$ diagrams, we observe that in the saturated phase, the $f$-mode undergoes significant broadening, consistent with previous studies \citep{2014ApJ...795L...8S}. To investigate this in more detail, we analyze the power spectrum at different values of $\tilde k_x$. Figure~\ref{fig:4} (right) presents a representative example. In the figure, we show the dimensionless spectral mode amplitude \(\tilde{P}_{\tilde{k}_x}(\tilde{\omega})\) at \(\tilde{k}_x = 2\) as a black dotted curve. Panel (a) corresponds to the kinematic phase, and panel (b) to the saturated phase. In both panels, three distinct peaks are visible, corresponding to the \(f\)-, \(p_0\)-, and \(p_1\)-modes from left to right. The fitted Lorentzian profiles are overplotted: blue for the \(f\)-mode, orange for the \(p_0\)-mode, and green for the \(p_1\)-mode. 

In the kinematic phase (panel (a) in Figure~\ref{fig:4}), each mode is fitted separately and shown here for completeness. In contrast, in the saturated phase (panel (b) in Figure~\ref{fig:4}), the \(f\)-mode broadens significantly and partially merges with the \(p_0\)- and \(p_1\)-mode, particularly at low \(\tilde{k}_x\). Therefore, all three modes are fitted together in this case. The combined fit is shown as a black dashed curve in panel (b), clearly capturing the broadened nature of the \(f\)-mode.

\begin{figure*}
  \centering
  \begin{minipage}[c]{0.48\textwidth}
    \centering
    \vspace{8mm}
    \includegraphics[width=\linewidth]{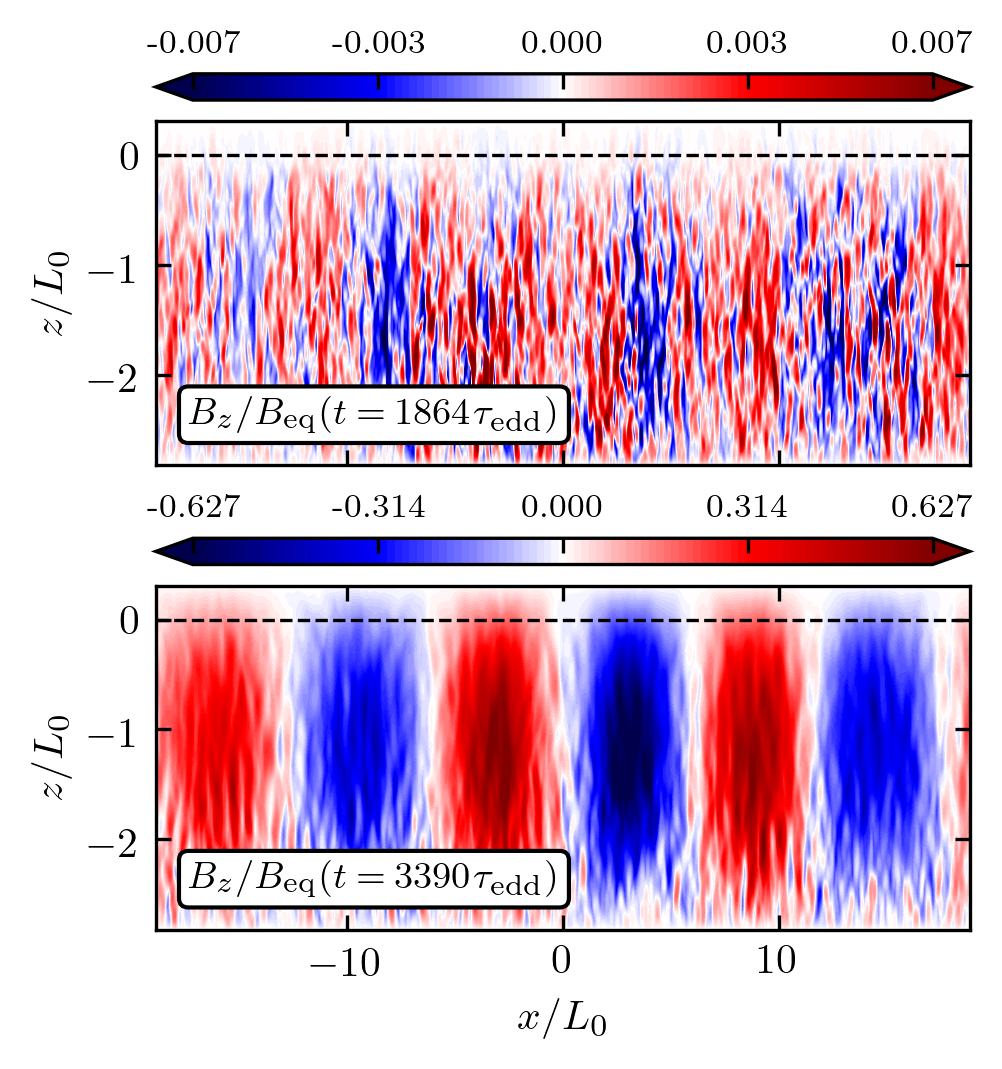}
  \end{minipage}
  \hfill
  \begin{minipage}[c]{0.48\textwidth}
    \centering
    \includegraphics[width=\linewidth]{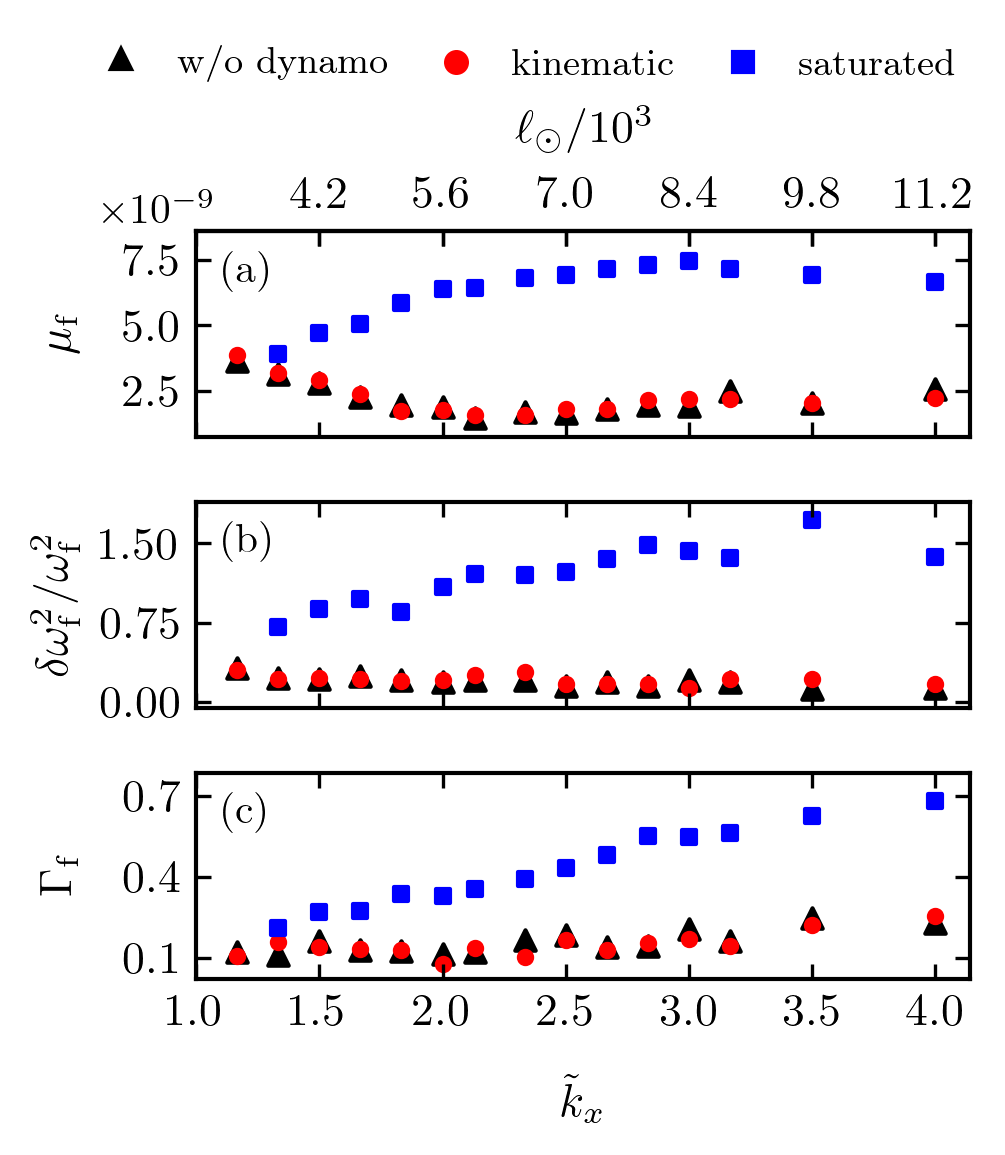}
  \end{minipage}
  \captionof{figure}{Left: normalized vertical magnetic field $B_z/B_{\rm eq}$ at the end of the \textit{kin} interval (top) and at the beginning of the \textit{sat} interval (bottom), as defined in the text. Right: Variation of different $f$-mode characteristics with $\tilde{k}_x$ for different runs. 
  (a) Mode strength ($\mu_f$), (b) relative frequency shift, and
  (c) full width at half-maximum (FWHM) of the $f$-mode, all as functions of $\tilde{k}_x$. 
  Tick marks at the top indicate the corresponding spherical harmonic degree $\ell$.}
  \label{fig:5}
\end{figure*}

To understand the physical origin of this change in the mode properties, we examine the structure of the magnetic field during the two phases.
In Figure~\ref{fig:5}(left) we show the normalized vertical magnetic field at two representative times: near the end of the \textit{kin} interval (top) and at the beginning of the \textit{sat} interval (bottom). A clear difference in the spatial structure of the magnetic field is evident. During the kinematic phase, the magnetic field remains relatively weak and lacks coherent large-scale organization. In contrast, in the saturated phase, the field develops a well-defined large-scale structure consistent with a Beltrami-type configuration, as expected for an $\alpha^2$ dynamo.

Note that, in both intervals, most of the magnetic field resides below the interface ($z=0$), which is indicated by the dashed black line.
We use perfect conductor boundary conditions in the vertical direction
for the magnetic fields, and therefore, the vertical component is suppressed
near the top boundary. Solar magnetic fields are not constrained in
this manner, and therefore, strong vertical magnetic fields are present
near the photosphere. Consequently, the magnetic configuration obtained
here should not be interpreted as a realistic model of the solar large-scale
dynamo. Rather, we use it to investigate the effects of dynamo-generated subsurface magnetic fields on the $f$-mode, which is expected to serve as an important diagnostic tool in predicting the emergence of the active regions at the solar surface.

Having characterized the differences between the magnetic field configurations in the two phases, we now quantify how these changes influence the properties of the $f$-mode.

From the fitted profiles of the $f$-mode for two different phases, we calculate three mode parameters prescribed in section \ref{sec:3.1} for various $\tilde{k}_x$ and for different runs. Figure~\ref{fig:5}.(right) summarizes our results. The parameters for the nonmagnetic case (h1) are shown by black triangles, while circular (red) and square markers (blue) denote the kinematic and saturated phases of the d1 run, respectively. 

From Figure~\ref{fig:5}(right), we observe that $f$-mode characteristics calculated from the nonmagnetic case (h1) and the kinematic phase of magnetic run d1 are identical. This behavior is evident in all three panels, indicating that the kinematic phase of a dynamo effectively acts as a nonmagnetic case. 

What is particularly striking, however, is the behavior of the $f$-mode in the saturated phase of the dynamo. Panel (a) shows a clear enhancement of the $f$-mode strength, indicating a strengthening of the $f$-mode. During the kinematic phase, the dynamo converts kinetic energy to magnetic energy, leading to the growth of the magnetic field. Once magnetic energy becomes comparable to the kinetic energy, the magnetic field begins to influence the flow through the Lorentz force. This feedback quenches the fluid flow, and through the nonlinear interaction, the magnetic field saturates to a finite amplitude. This quenching of the fluid flow results in a reduction of the system's kinetic energy. Naively, this reduction would suggest a corresponding decrease in the mode strength. However, our results reveal that the interaction between the dynamically evolving large-scale magnetic field and the $f$-mode instead leads to its strengthening. This enhancement is directly proportional to the magnitude of the magnetic field near the interface, where the $f$-mode is primarily localized.

In addition to the increase in mode strength, we also observe the other well-known effects of the magnetic field on the $f$-mode. Panel (b) shows that the magnetic field shifts the $f$-mode frequency, with the magnitude of the shift increasing with wavenumber. Panel (c) further demonstrates that the FWHM of the $f$-mode exhibits a similar trend. In the hydrodynamic case, we also observe $f$-mode broadening, which we attribute to the turbulent nature of the flow. The line widths in the kinematic phase of the d1 run (red circles) closely match those of the hydrodynamic case. In contrast, in the saturated phase, the broadening is significantly larger. Such broadening of the $f$-mode was first reported by \citet{2014ApJ...795L...8S}, who referred to it as a 'fanning out' of the $f$-mode and attributed it to the nonuniform nature of the magnetic field.

Throughout this section, our results have been presented in terms of the nondimensional wavenumber $\tilde{k}_x$. To assess their observational relevance, it is useful to relate $\tilde{k}_x$ with the spherical harmonic degree $\ell$. To do so, we recall that $\tilde k_x = k_x L_0$, where $L_0 = \gamma H_\mathrm{d}$. For the Sun, we approximate this by choosing $\gamma\sim 5/3$ and $H_\mathrm{d} = H_\odot \sim 0.15~\mathrm{Mm}$, which is roughly the scale height near the photosphere. Hence, the corresponding wavenumber for the Sun is given by $k \sim \tilde{k}_x / (\gamma H_\odot)$, where $\gamma H_\odot \sim 0.25~\mathrm{Mm}$. This leads to the relation $\ell \sim \tilde{k}_x R_\odot / (\gamma H_\odot) \sim 2800~\tilde{k}_x$. Using this relation,
the corresponding spherical harmonic degree $\ell$ for each value of $\tilde{k}_x$ is shown along the top axis of Figure~\ref{fig:5}(right). This mapping implies that the effects identified in this study are expected to manifest at relatively high $\ell$ compared to those accessible in current observations. However, we note that the
isothermal stratification of the bulk as adopted in this study is too idealistic, and this leads to an overlap of the $f$- and $p$-modes for $k_x\lesssim 1$. It is therefore likely that the effects seen in this work would continue to be relevant for lower $\ell$ in a more realistic polytropic bulk, which will be explored in a separate study.

From the $k$-$\omega$ diagrams shown in \Figs{fig:2}{fig:4}, and visually
comparing the $p$-mode ridges, we find that the $p$-modes too are relatively more broadened and their frequencies are higher when the magnetic fields saturate in the dynamo run as compared to the purely hydrodynamic case. Further investigations on the properties of $p$-modes in the presence of dynamo-generated magnetic fields will be taken up elsewhere.

\section{Conclusions}\label{sec:5}

Previous studies of this kind often relied on magnetic fields that are imposed artificially, with arbitrary amplitudes chosen a priori. Here, in this work, we relax this by including an $\alpha^2$ dynamo in the lower subdomain of the box, allowing the magnetic field to evolve on its own. For the $f$-mode properties,
we find that the kinematic phase of such a dynamo run is equivalent to a nonmagnetic case. All three $f$-mode parameters, namely the mode strength, frequency shift, and line width for the kinematic phase, are almost the same as those obtained from the nonmagnetic run h1. This is expected as, in the kinematic phase, the magnetic field is too weak to affect the velocity field; hence, the characteristics of the $f$-mode are the same in both cases. 

But as the dynamo evolves and generates large-scale magnetic fields, it begins to interact with the velocity field, thereby affecting the $f$-mode. These dynamo-generated magnetic
fields lie predominantly below the surface;
see Figure~\ref{fig:5}(left).
We find that the magnetic field not only increases the frequency of the $f$-mode, but also strengthens it and fans it out, which is clearly noticeable in the $k$-$\omega$ diagram; see Figure~\ref{fig:4}. This increase in the mode strength depends on the magnitude of the magnetic field near the surface, with these effects being
more pronounced at larger wavenumbers.

One possible caveat in this and other similar works with isothermal stratification
in the bulk is that the $f$- and $p$-modes overlap and interfere with each
other for $\tilde{k}_x \lesssim 1$, i.e., for $\ell \lesssim 2800$.
This limits our ability to explore the regime where current observations are more
sensitive. To address this, one needs to consider more realistic polytropic stratification
for the bulk. This will be taken up in a future study.

We emphasize that the present setup is not intended to model the solar dynamo. However, due to convective helical turbulence, one expects that the
$\alpha$-effect is operative in the Sun \citep{Singh_et_al_2018_Bihelical}. This is the motivation
for our choice of using helical forcing to enable the
$\alpha$-effect in our dynamo simulations, although solar magnetic fields are more complicated and modeling those is beyond the scope of this work.
The dynamo in our simulations serves only as a mechanism to generate a near-equipartition large-scale magnetic field self-consistently. This allows us to study the coupling between the $f$-mode and a dynamically generated magnetic field without imposing the field configuration artificially.

In this work, we have investigated the effects of dynamo-generated subsurface magnetic fields on the $f$-mode,
which shows strengthening as compared to the nonmagnetic case
or the weakly magnetized kinematic phase of the dynamo.
This is in agreement with the more idealized earlier works, which reported $f$-mode strengthening due to subsurface magnetic
fields.
We envisage that the $f$-mode, being sensitive to the magnetically induced
perturbations, may therefore serve as an important diagnostic tool in predicting the formation of the active regions on the solar surface.

\section{Acknowledgments}  
We thank the referee for many useful suggestions.
All simulations were performed on the Pegasus cluster at IUCAA. We sincerely acknowledge the computing facilities provided by IUCAA. We thank Kishore Gopalakrishnan for valuable discussions and insights.

\emph{Data Availability:}
Simulation setups and processed data used in this Letter are publicly available on Zenodo at DOI: \url{https://doi.org/10.5281/zenodo.20324776}.

\bibliography{mybib}{}
\bibliographystyle{aasjournal}

\end{document}